# Photonic Metamaterial Analogue of a Continuous Time Crystal

Tongjun Liu[1], Jun-Yu Ou[1], Kevin F. MacDonald[1], and Nikolay I. Zheludev[1,2]*

[1]*Optoelectronics Research Centre and Centre for Photonic Metamaterials, University of Southampton; Highfield, Southampton, SO17 1BJ, UK.*

[2]*Centre for Disruptive Photonic Technologies, School of Physical and Mathematical Sciences and The Photonics Institute, Nanyang Technological University, Singapore; 637378, Singapore.*

*Email: zheludev@soton.ac.uk

Time crystals are an eagerly sought phase of matter with broken time-translation symmetry. Quantum time crystals with discretely broken time-translation symmetry have been demonstrated in trapped ions, atoms and spins while continuously broken time-translation symmetry has been observed in an atomic condensate inside an optical cavity. Here we report that a classical metamaterial nanostructure, a two-dimensional array of plasmonic metamolecules supported on flexible nanowires, can be driven to a state possessing all of the key features of a continuous time crystal: continuous coherent illumination by light resonant with the metamolecules' plasmonic mode triggers a spontaneous phase transition to a superradiant-like state of transmissivity oscillations, resulting from many-body interactions among the metamolecules, characterized by long-range order in space and time. The phenomenon is of interest to the study of dynamic classical many-body states in the strongly correlated regime and applications in all-optical modulation, frequency conversion and timing.

Isotropic homogeneous matter is invariant under space-translation symmetry while in crystals with periodic atomic lattices that symmetry is broken. In nature, these symmetry-breaking crystalline states with long range order are achieved spontaneously through a phase transition (e.g. from water to ice). In recent years, the physics community has been captivated by the newly described phase of matter known as a "time crystal", with broken time-translation symmetry, analogous to conventional crystals in which space-translation symmetry is broken. A time crystal, as originally proposed by Wilczek[1], is a quantum many-body system whose lowest-energy state is one in which the particles are in continuous oscillatory motion. Although it has been shown that such a closed system, breaking continuous time-translation symmetry by exhibiting oscillatory dynamics, is prohibited by nature[2], a number of systems which show discrete time-translation symmetry-breaking imposed by an external modulated parametric drive have been realized on various platforms, including trapped atomic ions, Kerr-nonlinear microcavities, spin impurities, ultracold atoms, waveguide arrays, condensates of magnons, microwave systems and quantum computers[3-15].

Recently, a quantum time crystal that breaks time-translation symmetry continuously has been observed in an atomic Bose-Einstein condensate inside an optical cavity, manifested in the emergence of spontaneous oscillations of the intracavity photon number under optical pumping[16]. In this experiment, the pump laser was operated continuously, thereby respecting continuous time-translation symmetry while inducing a transition to symmetry-broken state. This atomic system realizes the spirit of Wilczek's original proposal more closely than discrete time crystals and could be a platform for exploring entanglement and topological phases in arrays of levitated nanoparticles and quantum simulations.

A defining characteristic of a continuous time crystal is a spontaneous transition, in a many-body system, to a robust oscillatory state in reaction to a time-independent external stimulus. Here we show that a two-dimensional periodic lattice of plasmonic metamolecules supported on doubly-clamped nanowire beams cut from a semiconductor membrane (Fig. 1) spontaneously transitions, at room temperature, to a state characterized by persistent optical transmissivity oscillations when illuminated by coherent light that stimulates interaction among the metamolecules. Above a threshold level of incident optical power, the spectrally dispersed thermal fluctuations of the individual nanowires become spatially coherent synchronous oscillations over the illuminated ensemble.

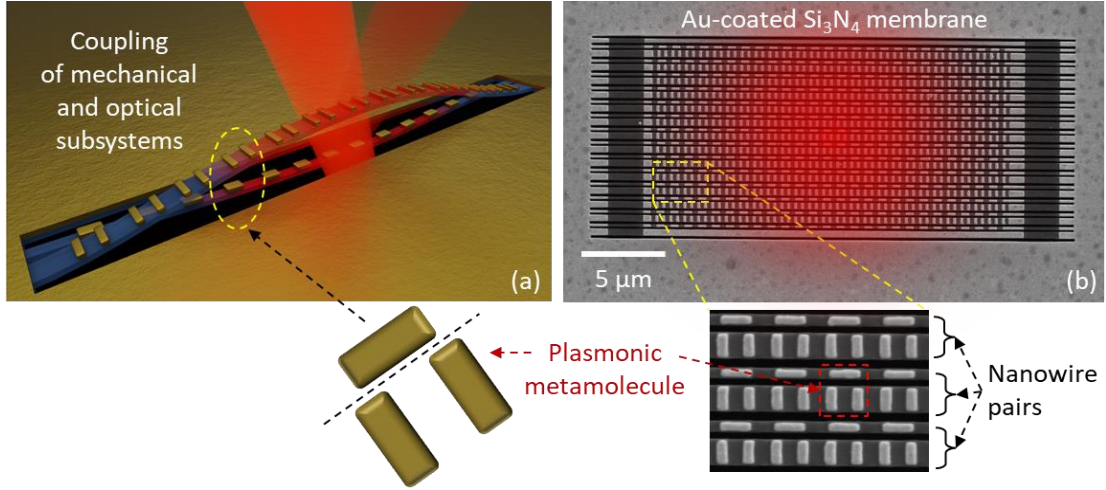

**Fig. 1. Photonic metamaterial analogue of a continuous time crystal** - A two-dimensional array of plasmonic metamolecules is supported by nanowires cut from a semiconductor membrane. (a) Artistic impression of the basic building block of the crystal – a pair of silicon nitride nanowires decorated with plasmonic [gold] metamolecules. (b) Scanning electron microscope image of the entire 2D array. Illumination with coherent light [as indicated by the schematically overlaid laser spot] induces a transition to a state of persistent synchronized nanowire oscillations.

Here it should be noted that the term 'photonic time crystal' bears some ambiguity: In the condensed matter physics and cold atom communities, and in the present work, it denotes a stable phase of matter that spontaneously breaks continuous time-translation symmetry through multi-body interactions. However, in the photonics community the terminology is often employed more broadly in reference to active systems relying on externally induced modulation of their constitutive parameters (see the review Ref. [17]).

The experimental sample was fabricated by focused ion beam milling from a gold-coated silicon nitride nano-membrane (Fig. 1b). The plasmonic metamolecules - Π-shaped arrangements of three gold nanorods - are supported on pairs of neighboring nanowires in such a way that the nanowires' mutual displacement (Fig. 1a) strongly affects the metamolecules' resonant plasmonic scattering properties. This coupling between mechanical and optical properties of the lattice provides opportunity for detecting the state of the lattice optically by monitoring its transmissivity. Indeed, it was shown recently that even picometric thermomechanical fluctuations in such structures can measurably modulate scattered and transmitted light[18].

In such an array, the nanowires can be described by set of $i$ coupled classical oscillators with the following equation of thermally-driven motion in the out-of-plane direction:

$$\ddot{x}_i + \Gamma_i \dot{x}_i + \omega_{0i}^2 x_i + \alpha x_i^2 + \beta x_i^3 + \sum \xi_{ij}(I)\,(x_i - x_j) = \sqrt{2\pi k_B T \Gamma_i / m_i}\,\eta(t) \quad (1)$$

According to the Langevin treatment of thermally driven oscillators, the right-hand side of the equation is the time-dependent thermal force experienced by the oscillator, which depends on the dissipation factor $\Gamma_i$ through the fluctuation-dissipation theorem. $\eta(t)$ is a normalized white noise term, $m_i$ is the effective mass of the nanowire, $k_B$ is the Boltzmann constant, $T$ is the temperature, and $\omega_{0i}$ ($\gg \Gamma_i$) is the nanowire's natural angular frequency of oscillation. A nanowire clamped at both ends exhibits 'geometric nonlinearity' (terms $\alpha x_i^2$ and $\beta x_i^3$) that becomes significant when the amplitude of oscillation approaches the thickness of the nanowire[19]. Due to fabrication imperfections and intrinsic variations of stress in the membrane in the present case, the fundamental frequencies of the nanowires $\omega_{0i}$ are dispersed within the range of about 2% around a central frequency $\omega_0 = 2\pi \times 0.99$ MHz.

It has been shown that in absence of noise, a pair of interacting nonlinear oscillators can be driven into the *discrete* time crystal state by parametric modulation of their natural frequencies[20]. Here we experimentally demonstrate a different route than leads to the *continuous* time crystal state: the combination of noise and interaction among oscillators induced by continuous illumination triggers a first-order dynamic phase transition to a state in which time-translation symmetry is broken. When the lattice of plasmonic metamolecules is illuminated, an optically controllable interaction between nanowires, described by the term $\xi_{ij}(I)(x_i - x_j)$ in the above equation of motion, emerges through dipole-dipole coupling of the induced plasmonic (gold nanorod) excitations. Here the coupling coefficient is a function of light intensity $I$ and the mutual position of the oscillators. The time-dependent stochastic thermal force on the right-hand side of the equation provides the seed noise needed for synchronization.

The array is illuminated with a single continuous beam of laser light at a wavelength of 1550 nm (close to the plasmonic absorption resonance of the metamolecules, see Supplementary materials), which is used both to monitor the transmissivity of the array as a function of time and to induce the synchronized oscillatory state. The sample is housed in a low vacuum optical cell with pressure maintained at $10^{-4}$ Torr to suppress damping of nanowire movements. Light is normally incident on the array, polarized parallel to the nanowires, and focused to a spot with a full-width half-maximum diameter of ~5 μm. The transmitted light intensity is monitored by a silicon photodetector.

At low laser power (≤tens of μW), the transmissivity spectrum contains several overlapping peaks of small amplitude at frequencies just below 1 MHz corresponding to oscillations of the individual illuminated nanowires. This is typical of an ensemble of nanowire oscillators with uncorrelated thermomechanical oscillation dynamics of characteristic amplitude $\langle x \rangle = \frac{1}{\omega_{0i}}\sqrt{\frac{k_B T}{m_i}}$ ~250 pm. While adiabatically increasing laser power, we initially observe red shifting of the spectral peaks, due to thermal expansion of the nanowires induced by laser illumination, as described in Ref. [21]. At around 130 μW, the onset of synchronization is observed in a narrowing of the spectrum. Further increasing the laser power leads to spontaneous synchronization of the nanowires' oscillation, manifested in the emergence of a single narrow peak in the transmissivity spectrum, with an amplitude four orders of magnitude larger than the peaks associated with the individual nanowires' uncorrelated thermomechanical fluctuations. While adiabatically decreasing laser power, we observe hysteretic loss of synchronization in reversion of the transmissivity spectrum to its low-power form (i.e. comprising several small amplitude peaks) below about 115 μW, revealing the first order nature of the synchronization phase transition.

One of the main features of a continuous time crystal is that, once initiated by an external stimulus, oscillations persist over arbitrarily long times, and this is true of our experimental system: Figure 2a shows that the spectrum of synchronized transmissivity oscillations is invariant as a function of time, for a constant level of incident optical power (ramped initially from zero to $P = 150$ μW, whereby a strongly synchronized state is achieved, and then held at $P = 124.3$ μW). The observed transmissivity oscillations are also stable against small perturbations: Fig. 2b shows slow (adiabatic) modulation of the transmissivity oscillation frequency driven by thermal expansion/contraction of the nanowires with modulation of incident power (and thereby nanowire temperature). Here, the incident light intensity is modulated by 5% at a frequency of 260 Hz, and the temperature of the nanowires adiabatically increases/decreases with laser power as their cooling time of ~85 μs is much shorter than the 3.85 ms period of intensity modulation. Here we can also see that the periodic modulation of transmissivity is weakly anharmonic - the transmitted light has a spectral component of modulation at the second harmonic of the main out-of-plane mode of the structure. This is explained by the geometric nonlinearity of nanowire motion: for nanowires of asymmetric cross-section (semiconductor beams decorated on one side with plasmonic nanorods), both quadratic and cubic terms are present in the equation of motion (terms $\alpha\, x_i^2$ and $\beta\, x_i^3$ in equation (1)).

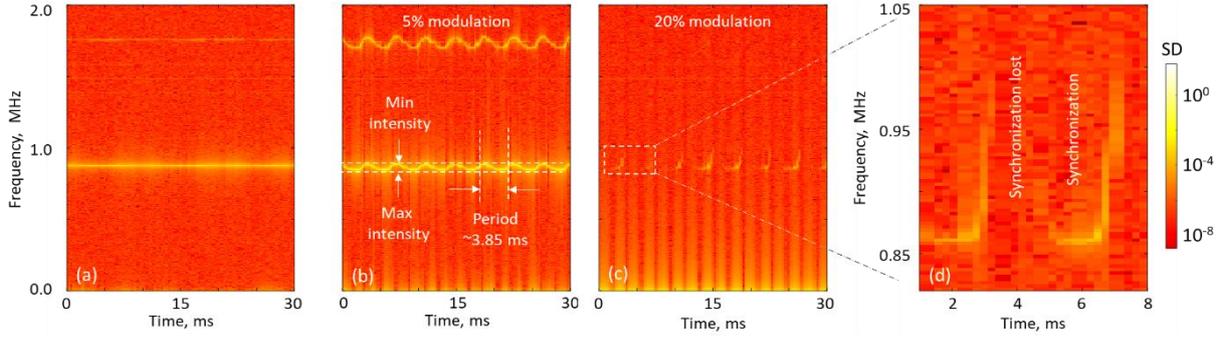

**Fig. 2. Persistence and stability and interruption of synchronized oscillations.** Evolution of the spectrum of transmissivity oscillations: (a) with time-invariant incident optical power above the threshold of synchronization, whereby synchronized oscillation persists indefinitely and the frequency of oscillation does not change; (b) with incident power sinusoidally modulated to a depth of 5% [from the $P$ = 124.3 µW level, at 260 Hz], whereby synchronized oscillation is maintained at a frequency that adiabatically dependents upon incident power; (c) at higher level of incident optical power modulation (20%), whereby synchronization is periodically suppressed and recovered, giving rise to transmissivity oscillation bursts, shown in closer detail in (d).

When the incident optical power is more strongly modulated (20%), periodic suppression and recovery (bursts) of transmissivity modulation are observed, as illustrated in Figs. 2c, d. In both cases (5 and 20% incident power modulation), periodic shifting of the nanowires' natural frequencies in response to modulation of the incident power leads to their synchronous oscillations being periodically chirped. As such there is a low frequency spectral component extending from zero to the chirp frequency, which is visible as a pattern of low frequency 'fringes' on the logarithmic scale of Figs. 2b and 2c. The fringes increase in brightness with increasing magnitude and rate of change of laser power, and disappear when the power reaches maxima and minima - hence two fringes are observed for each period incident laser power modulation.

To illustrate that synchronized oscillations have arbitrary phase each time a burst is generated, as is characteristic of a continuous time crystal[16], we analyzed their phase distribution: Figure 3a shows a typical time domain trace of such a burst. We find that the phase of oscillation at the dominant frequency, relative to the cycle of incident light modulation, is indeed randomly distributed over the [0, 2π] range, as illustrated in Fig 3b where the real and imaginary parts of the amplitude of the dominant spectral components are presented for 155 consecutive bursts.

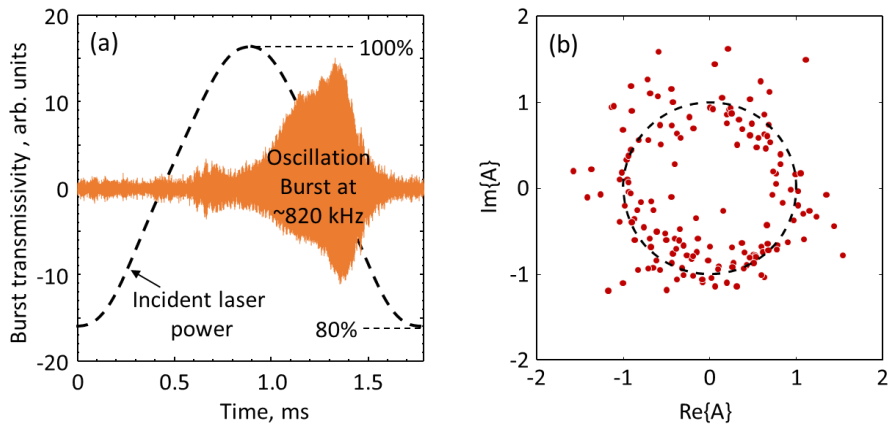

**Fig. 3. Synchronization phase.** (a) Time domain trace of a single transmissivity oscillation burst [orange line] overlayed on a trace showing the time dependence of incident laser power [in the 20% modulation regime, dashed black line]. (b) Imaginary vs. real part of signal amplitude at the fundamental oscillation frequency for 155 consecutive burst of synchronous transmissivity oscillation. [incident light intensity modulation at 560 Hz.].

Figure 4 presents analysis of the structure of a single spontaneous burst of oscillation resulting from a cycle of 20% decrease and recovery in incident laser power (as presented in Fig 2d). To evaluate how the peak frequency and amplitude of the burst evolves in time and with incident laser power we evaluated Fourier spectra of transmissivity for 25 consecutive time intervals over one incident laser power modulation period. This analysis yields values of the peak amplitude, width and frequency of oscillation spectral density (Fig 4a-c) at the average value of incident laser power for each interval.

We observe that in the dynamic regime of cyclically increasing and decreasing incident laser power, the spontaneous transition to and from synchronization has all of the characteristics of a first order phase transition, where the oscillation spectral density peak amplitude can be used as an order parameter for synchronization (here and below, terminology for states of the oscillator ensemble is adopted from Ref. [22]). With increasing laser power (orange symbols in Fig. 4) the array of oscillators is initially in an incoherent state (IS), characterized by small amplitude oscillations, wherein measurements of peak frequency and width are not possible. When the incident laser power exceeds ~130 µW, a weakly synchronized state (WSS) emerges, and the amplitude of oscillation spectral density grows exponentially, as indicated by the orange dashed line in Fig. 4a. At an incident laser power of ~150 µW, a strongly synchronized state (SSS) is achieved, which is characterized by the oscillation spectrum collapsing to a line of <2 kHz width. With decreasing power (blue symbols in Fig. 4a), the SSS state persists until the incident power falls below ~100 µW and the system returns to the IS, completing a full hysteresis loop as is characteristic of a first-order phase transition. Importantly, a well-defined stable frequency of oscillation ~855 kHz that is achieved in the SSS regime, which remains invariant and resilient to the variations in the level of light-induced heating until power drops to ~120 µW (Fig 4a, c), and is stable from burst to burst with variance of 1.115 kHz. The pronounced dependence of the frequency of transmissivity oscillation on incident power below 120 µW (Fig. 4c) is due to thermal contraction of the nanowires induced by decreasing laser illumination[18].

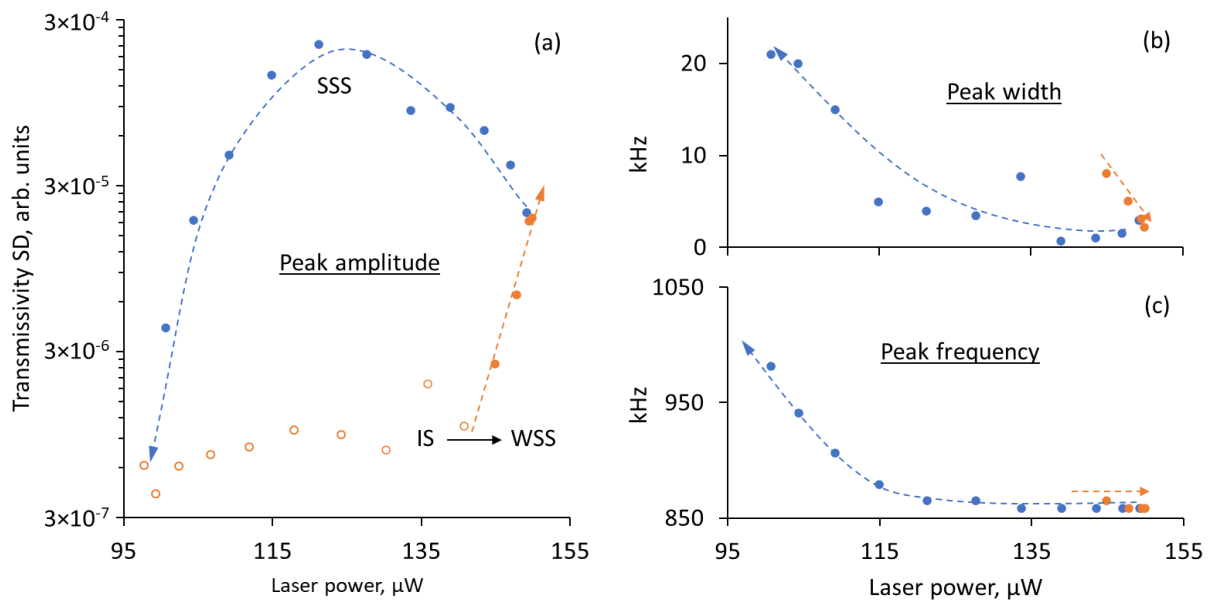

**Fig. 4. Synchronization as a first-order phase transition process.** Plates (a-c) present analysis of a burst of metamaterial array transmissivity oscillation during a single cycle of 20% incident laser power modulation. Plates show (a) the peak amplitude, (b) spectral width, and (c) central frequency of transmissivity spectral density for increasing (orange symbols) and decreasing (blue) incident laser power. Oscillatory regimes referred to as the incoherent state [IS], weakly synchronized state [WSS], and strongly synchronized state [SSS] are annotated in (a). For increasing laser powers from zero up to the IS-to-WSS transition, the small level of transmissivity modulation does not allow reliable measurements of oscillation frequency or peak width. As such, no symbols are plotted in (b) and (c) in this range, and hollow symbols are plotted in (a).

We argue that the underlying dynamics of the observed light-assisted phase transition to a persistent, high amplitude transmissivity oscillation regime, the main feature of a time crystal, can be interpreted as resulting from the synchronization of the ensemble of noisy thermal opto-mechanical nanowire oscillators via light-induced coupling $\xi_{ij}(I)$. Indeed, as introduced by Kuramoto[23], populations of interacting self-oscillators with distributed natural frequencies can exhibit remarkable transitions from a disordered state to an ordered one as the strength of interactions is varied and exceeds a certain threshold: a macroscopic number of elements begin to be mutually entrained with a common frequency. Within this general approach, Tanaka *et al.* have shown[22] that the synchronization of a periodically driven system of oscillators with inertia exhibits a first-order phase transition to synchronization from an incoherent state (IS) to a hysteresis loop between two macroscopic states, a weakly synchronized state (WSS) and a strongly coherent synchronized state (SSS). Later Hong *et al.* analysed this phenomenon with particular attention to the effects of noise[24]. They found that the coupling needed to achieve a phase transition increases, and the hysteresis collapses, when thermal noise coming into the system increases. Indeed, simple energy considerations show (see Supplementary Materials) that the laser power threshold for synchronization should increase with temperature. A regime of synchronization in an ensemble of globally coupled phase oscillators with distributed intrinsic oscillator frequencies and external independent noise forces – a system closer to the present experiment - has also been investigated, indicating a nonequilibrium transition between desynchronized and synchronised states and the existence of bistability[25, 26].

The components of the metamolecules - plasmonic particles located on different nanowires - behave as induced dipoles driven by the total optical field. If the particles are identical (as on alternate nanowires) and the induced dipoles are in phase, the interaction between them is reciprocal and $\xi_{ij}(I) = -\xi_{ji}(I)$. If the particles are different in size/geometry and phase (as on neighbouring nanowires), so are their resonant properties and induced dipoles. Indeed, as the light-induced interaction between particles relies upon the combination of a conservative gradient force and a nonconservative radiation pressure force, the interaction between two different particles is fundamentally nonreciprocal: $\xi_{ij}(I) \neq -\xi_{ji}(I)$. The *actio et reactio* equality is broken, leading to the appearance of a non-reciprocal interaction force in the equation of motion Eq.(1). The non-conservative contribution emerges from the radiation pressure induced by the scattered fields, which constantly pumps energy into the system, and thus cannot be derived from a Hamiltonian[27]. Following from earlier theoretical works by Sukhov *et al.*[28] and Karasek[29], such non-reciprocal light-induced interaction has recently been observed experimentally with optically trapped particles[27]. From the theory of coupled oscillators, it is well known that nonreciprocal coupling does not split resonant frequencies but does lead to an increase in oscillation amplitude. We therefore argue that nonreciprocal coupling may play a key role in the synchronization of ensembles of oscillators.

Our ensemble of nanowire oscillators exhibiting thermal motion and interacting with help of an external light field can be seen as a classical analogue of an ensemble of quantum emitters interacting with a common light field and exhibiting Dicke superradiance[30]: If the wavelength of the light is much greater than the separation of the emitters, then the emitters interact with the light collectively and coherently emit light with intensity proportional to $N^2$ (superradiance), very differently from the emission with intensity proportional to $N$ observed for a group of independent atoms. Indeed, the relationship between the Dicke superradiant state and discrete time crystal ordering was recently theoretically generalized in terms of the Landau theory of phase transitions[31] and was also noted in regard to the recently demonstrated quantum continuous time crystal[16].

In summary, we have demonstrated an artificial photonic material - a two-dimensional array of plasmonic metamolecules supported on flexible nanowires - that can be driven to a state exhibiting all the defining characteristics of a continuous time crystal - a novel state of matter that continuously breaks time-translation symmetry: i) Time-independent excitation, in the form of illumination with coherent light that is resonant with the plasmonic modes of the metamolecules, spontaneously triggers strong periodic oscillation of optical transmissivity; ii) The oscillations result from many-body interactions among numerous metamolecules coupled by electromagnetic dipole-dipole interactions (induced by the incident light), whereby the individual thermal oscillations of nanowires are replaced by coherent, superradiant-like thermal motion of the illuminated ensemble; iii) The transition to the coherent oscillatory regime has the nature of a first-order phase transition; iv) The transmissivity oscillations are robust to small perturbations and the phase of the oscillation is random for different realizations; v) The state exhibits long range order in space, as manifested in synchronization of the illuminated ensemble; and long range order in time, as seen in the robust indefinite persistence of synchronous oscillation.

Apart from its intrinsic conceptual interest, our results, and the simplicity and control achieved with our nano-opto-mechanical platform, offer a new class of dynamical many-body system that expands the concepts of long-range order and spontaneous symmetry breaking into the time domain. It paves the way toward a comprehensive study of dynamic many-body states in the strongly correlated regime in a classical system that complements the cold atom and spin platforms where many-body quantum states of bosonic or fermionic matter can be investigated. These studies can be facilitated by direct access to the nanowire positions using recently developed atomic scale techniques for detecting localization and movement of the nanoscale strings and cantilevers[32][33]. The dynamics of transition from the stochastic and ballistic thermal motion of individual nanowires into the collective superradiant state at difference time scales and at various ambient temperatures is of interest for nanoscale thermodynamics[32].

Micro- and nanomechanical metamaterials have been demonstrated across the electromagnetic spectrum from microwave to terahertz and optical frequencies using a variety of materials, from silicon to high-index dielectrics and plasmonic metals, using focused ion beam milling, optical and electron beam lithography. The design of metamolecules affects the quality factors of resonances and bandwidth of such devices and as such the platform can be readily adapted to different applications. Step-like transitions to a robust superradiant mode induced by light may be used in optically controlled modulators and frequency mixers not yet known in photonic technologies; Such triggering could also be employed in novel all-optical timing applications where periods of oscillation are counted after triggering; And the random phase of spontaneous oscillations may be exploited in as yet undescribed all-optical pseudo-random number generators based on a classical platform, in contrast to generators exploiting the intrinsic randomness of quantum mechanics.

**Acknowledgements**

Authors thank Yidong Chong for critical comments that helped to improve the manuscript and Arkady Pikovsky for suggesting references on prior theoretical works on synchronization. This work was supported by the UK Engineering and Physical Sciences Research Council (grants EP/M009122/1), the Singapore Ministry of Education (grant MOE2016-T3-1-006) and the China Scholarship Council (TL, grant 201806160012).

# Supplementary Material:
# Photonic Metamaterial Analogue of a Continuous Time Crystal

Tongjun Liu[1], Jun-Yu Ou[1], Kevin F. MacDonald[1], and Nikolay I. Zheludev[1,2]*

[1]*Optoelectronics Research Centre and Centre for Photonic Metamaterials, University of Southampton; Highfield, Southampton, SO17 1BJ, UK.*

[2]*Centre for Disruptive Photonic Technologies, School of Physical and Mathematical Sciences and The Photonics Institute, Nanyang Technological University, Singapore; 637378, Singapore.*

*\*Email: niz@orc.soton.ac.uk*

## Section 1: Metamaterial geometry, fabrication, and optical properties

The unit cell of the plasmonic resonator array contains a gold Π-shaped metamolecule formed by three gold nanorods supported on a two (geometrically different) silicon nitride nanowires [mechanical resonators], as illustrated in Fig. S1a.

The structure was fabricated by focused ion beam milling on a 50 nm thick, 500 × 500 μm$^2$, low stress (<250 MPa) silicon nitride membrane (Norcada Inc.) coated with 50 nm of thermally evaporated gold. First, the metamolecule (optical resonator) array was defined in the gold layer and the supporting nanowires (mechanical oscillators) were formed by milling through the silicon nitride membrane. Optical properties of the metamolecule ensemble are shown in Fig. S1b.

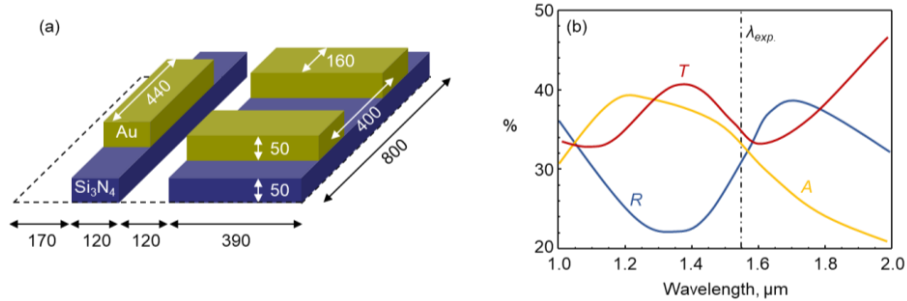

**Fig. S1: Optomechanical metamaterial – geometry and optical properties.** (a) Dimensional schematic of a metamolecule unit cell [in nm]. (b) Measured reflectivity *R*, transmissivity *T* and absorption *A* spectra of the metamaterial ensemble, measured using a microspectrophotometer with light incident on the gold side of the sample and polarized parallel to the nanowires [as in synchronization experiments]. The dot-dashed vertical line indicates the 1550 nm wavelength of laser illumination employed in synchronization experiments.

## Section 2: Light-induced interaction between plasmonic metamolecules

A detailed analysis of the synchronization process, requiring numerical modelling based upon the set of nonlinear equations (1) given in the main text of the paper, lies outside of the scope of the present study. Here, we present some qualitative estimates on the conditions leading to light-induced synchronization:

Each of 375 metamolecules in the array consists of two components: a single gold nanorod on a narrow silicon nitride nanowire with its long axis parallel to that of the wire (component A), and a pair of gold nanorods on a wide nanowire in the perpendicular orientation (component B), as illustrated in Fig. S1a. Within the metamolecule, the energy of light-induced interaction between its components can be evaluated as $V(r) = -\frac{Re\{\mu_A \mu_B\}}{2\pi\epsilon_0 \, r_{AB}^3}$, where $r_{AB}$ = 375 nm is the distance between the nanowires, $\mu_A$ and $\mu_B$ are the light-induced dipole moments of components A and B respectively, and $\epsilon_0 = 8.85\times10^{-12}$ F/m is the permeability of free space.

From numerical simulations (Fig. S2) using the method described in Ref. [1], the prevailing dipole moments parallel to the nanowires, induced by incident light at a wavelength of 1550 nm polarized in that directions with an incident power of 1 µW per unit cell (with area 0.64 µm$^2$, giving a power density $I$ = 1.56 µW/µm$^2$) are: $\mu_A$= 1.6×10$^{-26}$ + $i$ 9.1×10$^{-26}$ C·m and $\mu_B$= -3.2×10$^{-26}$ - $i$ 8.4×10$^{-26}$ C·m. This gives an intramolecular interaction energy of $V$ ~ 2.43×10$^{-21}$ J. Interactions between neighboring metamolecules are also dominated by the interaction between dipoles $\mu_A$ and $\mu_B$, but with slightly larger separation $r'_{AB}$ = 425 nm, giving an intermolecular interaction energy of $V'$ ~1.67×10$^{-21}$ J.

At the onset of synchronization, with a total laser power of ~110 µW incident on the array, the energy of dipole-dipole interactions between two neighboring nanowires (each supporting components of 25 metamolecules) is $V_{total}$ ~1×10$^{-19}$ J. A relative stochastic thermal out-of-plane displacement $x_r$ between two neighbouring nanowires will reduce the intramolecular interaction energy by $\delta V_{total} \approx \frac{3V_{total}}{2} \times \left(\frac{x_r}{r_{AB}}\right)^2$ and this is what drives the synchronization of their oscillations. Indeed, a small input of energy $\delta E$, related to movement in the out-of-plane direction, added to the potential energy $H$ of the oscillator will shift its oscillation frequency by $\delta\omega_0 = \frac{\delta E}{2H} \times \omega_0$. In the regime of thermal motion, equipartition dictates that $H = \frac{k_B T}{2}$ and therefore $\delta E = k_B T \frac{\delta\omega_0}{\omega_0}$. With the natural frequencies of nanowire oscillation being spread over a range $\delta\omega = 2\pi\times 6$ kHz around a central frequency $\omega_0 = 2\pi \times 0.94$ MHz, the additional energy of light-induced dipole-dipole interaction needed to initiate synchronization can be estimated as $V_s \sim \frac{k_B T}{2} \times \frac{\delta\omega}{n\,\omega_0} = 1.1\times10^{-24}$ J (assuming the spectral distributions includes six illuminated nanowire pairs ($n$ =12) within the laser spot). $\delta V_{total}$ becomes comparable to $V_s$ when the magnitude of relative out-of-plane displacement $x_r$ approaches ~1000 pm – a value that is consistent with the estimated ~380 and ~250 pm root-mean square amplitudes of stochastic thermal displacement of individual type A and B nanowires, respectively. Indeed, $\delta V_{total}$ here represents a lower estimate, as it does not account for interactions beyond nearest neighboring nanowires or higher multipole effects. As such, $x_r$ ~1000 pm is an upper estimate on the required amplitude of relative thermal displacement. We therefore conclude that light-induced dipole-dipole interactions between the components of metamolecules on neighboring nanowires, in the presence of thermal noise, are the main driver of nanowire synchronization.

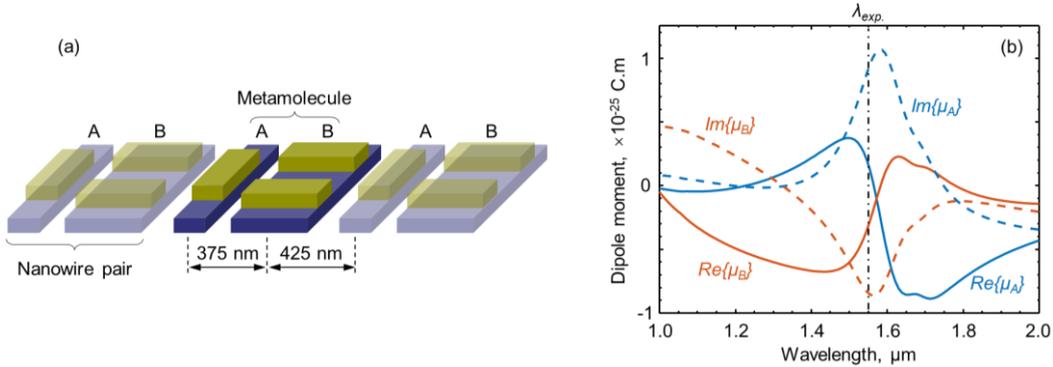

**Fig. S2**. **Light-induced inter- and intramolecular interactions.** (a) Schematic of metamolecule arrangement on neighboring nanowire pairs. (b) Dispersion of the dominant components of the real and imaginary parts of the dipole moments induced by incident light polarized parallel to the nanowires at an intensity of 1.56 µW/µm$^2$ [as labelled]. The dot-dashed vertical line labelled $\lambda_{exp.}$ indicates the 1550 nm wavelength of laser illumination employed in synchronization experiments.

2